\documentclass[sigconf]{acmart}
\usepackage{multirow}
\usepackage{bm}
\usepackage{arydshln}
\usepackage{tabularx}
\AtBeginDocument{%
  \providecommand\BibTeX{{%
    \normalfont B\kern-0.5em{\scshape i\kern-0.25em b}\kern-0.8em\TeX}}}

\copyrightyear{2023}
\acmYear{2023}
\setcopyright{acmlicensed}\acmConference[MM '23]{Proceedings of the 31st ACM International Conference on Multimedia}{October 29-November 3, 2023}{Ottawa, ON, Canada}
\acmBooktitle{Proceedings of the 31st ACM International Conference on Multimedia (MM '23), October 29-November 3, 2023, Ottawa, ON, Canada}
\acmPrice{15.00}
\acmDOI{10.1145/3581783.3612086}
\acmISBN{979-8-4007-0108-5/23/10}

\begin{document}

%%
%% The "title" command has an optional parameter,
%% allowing the author to define a "short title" to be used in page headers.
\title{Enhancing Fake News Detection in Social Media via Label Propagation on Cross-modal Tweet Graph}

%%
%% The "author" command and its associated commands are used to define
%% the authors and their affiliations.
%% Of note is the shared affiliation of the first two authors, and the
%% "authornote" and "authornotemark" commands
%% used to denote shared contribution to the research.
%\author{Anonymous}
\author{Wanqing Zhao}
\affiliation{%
  \institution{Northwest University}
  \country{China}}
\email{zhaowq@nwu.edu.cn}

\author{Yuta Nakashima}
\affiliation{%
  \institution{Osaka University}
  \country{Japan}}
\email{n-yuta@ids.osaka-u.ac.jp}

\author{Haiyuan Chen}
\affiliation{%
  \institution{Northwest University}
  \country{China}}
\email{chenhaiyuan@stumail.nwu.edu.cn }

\author{Noboru Babaguchi}
\affiliation{%
  \institution{Osaka University}
  \country{Japan}}
\email{babaguchi@ids.osaka-u.ac.jp}

%删除reference信息
%\settopmatter{printacmref=false} 
%\renewcommand\footnotetextcopyrightpermission[1]{}

%%
%% By default, the full list of authors will be used in the page
%% headers. Often, this list is too long, and will overlap
%% other information printed in the page headers. This command allows
%% the author to define a more concise list
%% of authors' names for this purpose.

%%
%% The abstract is a short summary of the work to be presented in the
%% article.
\begin{abstract}
Fake news detection in social media has become increasingly important due to the rapid proliferation of personal media channels and the consequential dissemination of misleading information. Existing methods, which primarily rely on multimodal features and graph-based techniques, have shown promising performance in detecting fake news. However, they still face a limitation, i.e., sparsity in graph connections, which hinders capturing possible interactions among tweets. This challenge has motivated us to explore a novel method that densifies the graph's connectivity to capture denser interaction better. Our method constructs a cross-modal tweet graph using CLIP, which encodes images and text into a unified space, allowing us to extract potential connections based on similarities in text and images. We then design a Feature Contextualization Network with Label Propagation (FCN-LP) to model the interaction among tweets as well as positive or negative correlations between predicted labels of connected tweets. The propagated labels from the graph are weighted and aggregated for the final detection. To enhance the model's generalization ability to unseen events, we introduce a domain generalization loss that ensures consistent features between tweets on seen and unseen events. We use three publicly available fake news datasets, Twitter, PHEME, and Weibo, for evaluation. Our method consistently improves the performance over the state-of-the-art methods on all benchmark datasets and effectively demonstrates its aptitude for generalizing fake news detection in social media.
\end{abstract}

%%
%% The code below is generated by the tool at http://dl.acm.org/ccs.cfm.
%% Please copy and paste the code instead of the example below.
%%

%%
%% Keywords. The author(s) should pick words that accurately describe
%% the work being presented. Separate the keywords with commas.

%%
%% This command processes the author and affiliation and title
%% information and builds the first part of the formatted document.
\maketitle

\section{Introduction}

The proliferation of social media has given rise to an abundance of non-official personal media channels that allow for the dissemination of news and opinions. Due to the absence of rigorous editorial oversight, such brand-new media has engendered a concomitant dissemination of fake news, which poses the potential to mislead readers and carry deleterious social consequences. Fake news may manipulate images and texts from various sources; generative AI \cite{cao2023comprehensive} even allows the generation of fake news from scratch by fabricating high-quality and specious images and text \cite{brown2020language,ramesh2021zero}, making it hard for readers to distinguish them from genuine ones.

\begin{figure}
\includegraphics[width=\columnwidth]{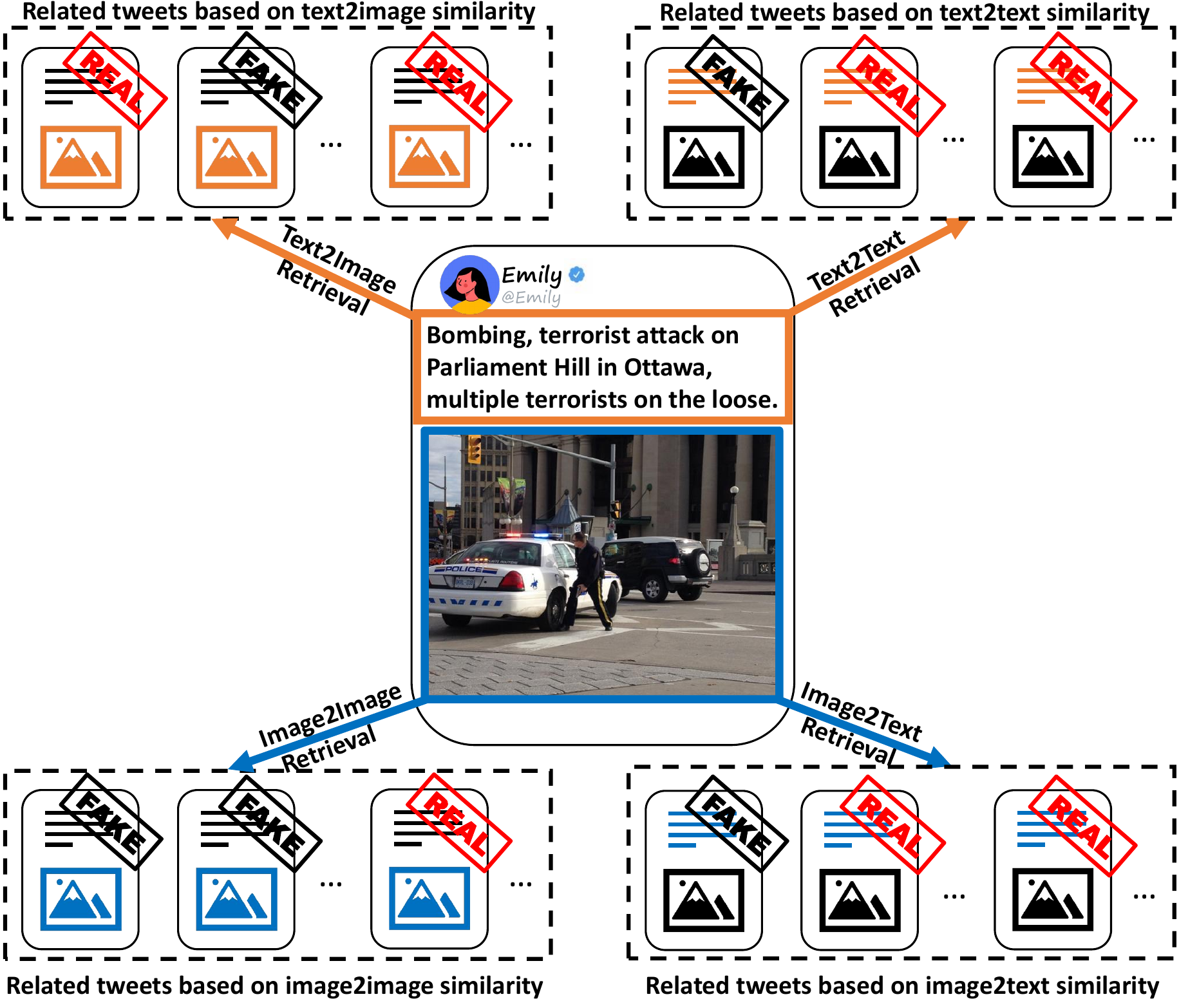}
\centering
\caption{\label{fig:intro} By utilizing cross-modal search, we can construct a more comprehensive connection to relevant tweets that may contain potential positive and negative correlations. These connections can be exploited to enable more credible detection of fake news.}
\end{figure}

Recent widespread access to new technologies makes automated detection of fake news imperative. To effectively identify fake news and mitigate its adverse effects, it is crucial to incorporate multimodal information, including both text and images. This allows for a more comprehensive analysis of social media tweets to improve the reliability of fake news detection. Several multimodal fake news detection approaches have been proposed, which fuse features from different modalities to achieve more accurate detection than text-only approaches \cite{singhal2019spotfake,qian2021hierarchical,zhou2020similarity}. Even with the fusion of images and text for detection, relying on a single tweet to determine the veracity of the original event can be one-sided and unreliable.

Some works have attempted to leverage explicit social context to build links between tweets, e.g., building a graph based on retweets and hashtags on tweets and using Graph Neural Networks (GNNs) \cite{zhou2020graph} to model the relationships among them \cite{yang2021rumor,wei2021towards,zhengmfan}. With the help of GNNs, neighboring tweets can be aggregated to obtain a more comprehensive representation. 

Graph construction solely based on social context may still be sparse. For example, multiple users can author an original tweet message with their own images on a single event, so their tweets cannot be in the tweet-retweet relationship. Also, the users do not necessarily use the same hashtags. Consequently, tweets on a single event may not always be connected to each other based on the social context. GNNs typically model interactions only between connecting tweets, and such sparse connections hinder capturing possible interactions among tweets. This sparsity can be problematic because the veracity of a certain tweet may stand only when it is compared with others. This observation inspires us to densify the connection between tweets based on the content of tweets. 

An impactful vision-and-language model, CLIP \cite{radford2021learning}, has recently been proposed to encode images and text into a unified space. 
%CLIP employs contrastive learning to ensure that paired images and texts have similar representations. CLIP is trained on a large corpus of 400 million noisy image-text pairs collected from the internet. 
CLIP's strong zero-shot capabilities for several tasks inspire us to leverage it for connecting related tweets in social media based on similarities in their text and images to densify the graph. As depicted in Figure~\ref{fig:intro}, a tweet can be utilized to retrieve relevant tweets using features obtained from CLIP through image-to-image or image-to-text search. Similarly, text-to-text or text-to-image search can be performed. We can then add an edge between a pair of relevant tweets. Such a densified graph capture interactions among more tweets by contextualizing features with connecting tweets.  

One interesting extension of this idea is to use the same graph for label propagation. Predictions for a pair of connecting tweets can have a certain correlation as their content is similar. This correlation, however, can be both positive and negative. As shown in Figure \ref{fig:intro}, a pair of semantically similar tweets can have opposite labels as real and fake news on the same event can share most of their content (e.g., tweets may come with very similar images of the same scene but their text messages can be different). Therefore, label propagation over our densified graph should account for this.

Based on the above motivations, we propose a method for detecting fake news on social media by leveraging a cross-modal tweet graph to model deeper interaction behind different tweets. We construct a cross-modal tweet graph by utilizing CLIP to extract potential connections based on similarities in text and images. We then design a Feature Contextualization Network with Label Propagation (FCN-LP). Feature Contextualization Network (FCN) learns the interaction between tweets over the graph, while Label Propagation (LP) 
incorporate the contextualized features to model both positive and negative correlations of predicted labels by assigning signed attention to edges. The propagated labels from the graph are then weighted and aggregated for the final detection. To improve the model's generalization performance, we introduce a domain generalization loss that ensures consistent features between seen and unseen tweets. 

\textbf{Contributions}. Our experiments show consistent performance improvement with FCN-LP over the state-of-the-art methods over three different fake news detection datasets. This means that CLIP-based graph construction successfully captures underlying structures of real and fake news tweets of unseen events. FCN further enhances the separation between real and fake news tweets. This implies that the FCN features encode the difference among them. We qualitatively confirm that LP can model positive and negative correlations from the FCN features. Our domain generalization loss is designed to match the distribution of tweets on unseen tweets (i.e., ones not used for training FCN) to the distribution of seen tweets. Given the performance boost, we would say that this loss consequently leads to features invariant to events. 

\begin{figure*}[ht]
\includegraphics[width=0.93\textwidth]{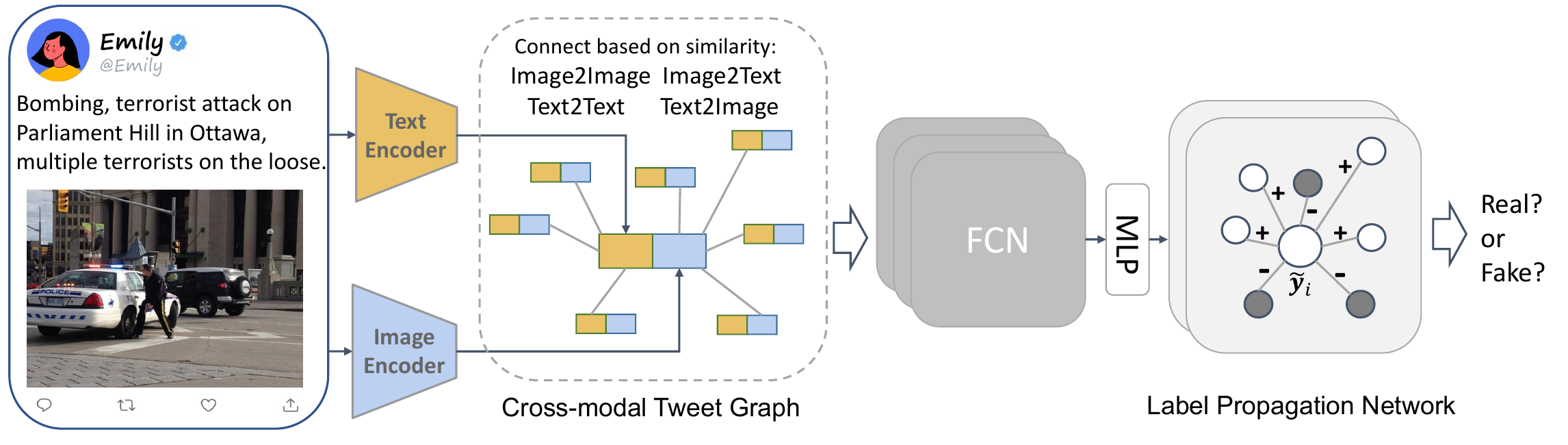}
\centering
\caption{\label{fig:overview}An overview of FCN-LP for fake news detection.} 
\end{figure*}

\section{Related Works}
%This section provides a short overview of the literature for detecting fake news. We then summarize the more relevant multi-modal and graph-based methods that are most relevant to our work, while discussing the differences between them.

%\textbf{Fake News Detection.} 
Fake news detection is often treated as a binary classification problem. Earlier approaches used features extracted from text content to train a fake news classifier \cite{castillo2011information,popat2017assessing}. These methods require expensive pre-processing and tricky feature engineering. Deep neural networks (DNNs) can build an end-to-end model from the original text to the classifier \cite{yu2017convolutional,shu2017fake}. 
Social network posts often consist of multimodal data, such as images and text, to make the post more informative. This abundance of data inspires researchers to propose multimodal approaches \cite{singhal2019spotfake,qian2021hierarchical,zhou2020similarity}, providing more comprehensive classification results. Moreover, some graph-based fake news detection methods explore social relationships to further enhance the representation of tweets \cite{yang2021rumor,wei2021towards,zhengmfan}. This section provides an overview of these methods. 

\textbf{Multimodal Fake News Detection.} In social media, tweets comprise not only text but are often associated with images and videos. Such multimodality provides complementary perspectives on the content of the tweet. As such, combining text and visual information can lead to a more complete understanding of the content of the tweet. This has motivated researchers to develop multimodal fake news detection. The most common formulation is to extract textual and visual features separately using DNNs and fuse the extracted features to classify the tweet into fake or real. 

Spotfake \cite{singhal2019spotfake} employs VGG19 \cite{simonyan2014very} and BERT \cite{kenton2019bert} to extract visual and textual features, respectively. These features are concatenated and fed into a classifier to leverage both modalities. HMCAN \cite{qian2021hierarchical} feeds image and text features into a multimodal contextual attention network. This approach captures both inter- and intra-modality relationships for a better fusion of the two modalities. Some studies compared the visual and textual content of news articles. SAFE \cite{zhou2020similarity} leverages an image-to-text model to convert visual information into textual descriptions, which enables a direct comparison between the two modalities. The representations of textual and visual information, along with their similarity, are jointly learned and used to predict fake news. 

All these methods treat tweets individually, i.e., they take a single tweet to make a prediction; they ignore the social context that can help fake news detection. For example, how the tweet spread (who retweeted or replied) can be a cue to identify fake news.

\textbf{Graph-based Fake News Detection.} Graph-based approaches have been proposed to explore the social context of a tweet as a graph structure to render the social context of fake news \cite{wei2021towards,zhengmfan,nguyen2020fang}. These methods evaluate the tweet from a holistic perspective, taking into account the relationships and interactions among different users (tweet authors) and sources. These approaches effectively capture the complex network of social relationships that contribute to the dissemination and credibility of a tweet. Bi-GCN \cite{wei2021towards} employs a Graph Convolutional Network (GCN) to organize the spread of fake news in social media. The method utilizes a top-down GCN to calculate the propagation of fake news and a bottom-up GCN to aggregate dispersed fake news. More relevantly, MFAN \cite{zhengmfan} sought to integrate textual, visual, and social graph features within a unified framework. 

One interesting extension of such graph-based approaches is to further incorporate the relationships and interactions among tweets on the same event. Social media typically have many tweets about a specific event, some of which can be fake. Without the knowledge of the event, fake news may only be spotted in the context of others on the same event; therefore, contextualizing a tweet may allow for identifying fake news tweets even if they do not stand by themselves. 

Inspired by this assumption, we construct a cross-modal tweet graph using the cross-modality of CLIP \cite{radford2021learning}, capturing potential connections between tweets with similar text or images. Based on this cross-modal tweet graph, a GCN with Label Propagation considers both positive and negative correlations between neighboring tweets, effectively integrating their corresponding prediction labels for a more accurate final decision.

\section{Method}
We primarily focus on utilizing a multimodal tweet (i.e., text and an image) and the corresponding label on the veracity of its content for training a model for fake news detection. We initially extract potential connections among tweets by leveraging their similarity to construct a cross-modal tweet graph. LPN then learns the signed weight of edges based on node features. This weight encodes positive or negative correlations between a pair of tweets to facilitate propagating their labels. %Finally, the propagated labels from the graph are then weighted and aggregated for the final detection. 
For training, we designed a domain generalization loss, asides from the cross-entropy loss, to improve the model's ability to detect unseen events. Figure \ref{fig:overview} shows the overall architecture of our model.
 
\subsection{Cross-modal Tweet Graph Construction}

As aforementioned, contextualizing a tweet with relevant ones (i.e., tweets on the same event) may facilitate the detection of fake news. A tweet is often associated with some tags that identify the corresponding event, but they are not always available and noisy. Instead of such unreliable sources, we use inter- and intra-modal similarities to construct a graph that potentially captures the relevance among tweets. 

Our method uses CLIP \cite{radford2021learning}, which encodes images and text into a unified space, for similarity computation. Formally, let $\mathcal{D}$ be a set of $|\mathcal{D}|$ tweets, where each tweet $d_i = (v_i, t_i) \in \mathcal{D}$ contains an image $v_i$ and text $t_i$. We also denote an undirected graph to represent the relevance among tweets by $\mathcal{G} = (\mathcal{D}, \mathcal{E})$. $\mathcal{E}$ contains edge $e_{ij} = (d_i, d_j)$ if the similarity between $d_i$ and $d_j$ is high. Specifically, for pair $(d_i, d_j)$, we extract CLIP features from both the visual and textual modalities and compute the cosine similarities for all possible combinations of modalities, denoted by $s_{ij}^c$, where $c$ is in $\mathcal{C}=\{\text{image-to-image}, \text{text-to-text}, \text{image-to-text}, \text{text-to-image}\}$. 
We define a similarity threshold, denoted as $\tau$, for all possible combinations within the set $\mathcal{C}$. A pair $e_{ij}$ is included in $\mathcal{E}$ if its corresponding similarity $s_{ij}^c$ surpasses $\tau$, i.e., $e_{ij} \in \mathcal{E}$ when $s_{ij}^c \ge \tau$ for at least one combination $c$.
%With $\text{Top-}K^c$ be a set of the pairs with the top-$K$ similarity for combination $c$, $e_{ij}$ is in $\mathcal{E}$ when $s_{ij}^c \in \text{Top-}K^c$. 

%\begin{equation}
%e_{ij}=
%\begin{cases}
%1, & \text{if } (I_i,I_j), (T_i,T_j), (I_i,T_j), (T_i,I_j) \in \text{Top-K}(S_{clip}); \\
%0, & \text{otherwise}.
%\end{cases} 
%\label{eq:edge}
%\end{equation}

Connected tweets in this graph have potential relevance; however, it is important to emphasize that their claims may differ. For example, two tweets may have the same image and are related to the same event but possibly say the opposite. It is also possible that the text of one tweet highly matches the image of another tweet, but their texts exhibit significant differences. These connections can detect fake news, but a well-designed graph-based model is needed to discern such differences.

\subsection{Feature Contextualization}

Feature contextualization network (FCN) aggregates relevant (connected) tweets for a certain tweet to contrast it with the others. Specifically, we employ Graph Convolutional Network (GCN)-based feature update to tweet feature $\mathbf{x}_i^l \in \mathbb{R}^d$ in the $l$-th layer ($l=1,\dots,L-1$), which can be written as:
\begin{equation}
\mathbf{x}_i^{l+1} = \sigma \left(W \mathbf{x}_i^l + \sum_{j \in \mathcal{N}_i} \frac{W^l \mathbf{x}_j^{l}}{\sqrt{|\mathcal{N}_i| \; |\mathcal{N}_j|}} \right),
\label{eq:gcn}
\end{equation}
where $\sigma(\cdot)$ is an activation function (e.g., ReLU), $W^l$ is a learnable weight matrix, and $\mathcal{N}_i$ is the set of connected tweets (nodes) to $d_i$, given by $\mathcal{N}_i = \{d_j | (d_i, d_j) \in \mathcal{E}\}$. We normalize the aggregated neighbor tweets' features by the square root of the product of the degrees $|\mathcal{N}_i|$ and $|\mathcal{N}_j|$. The input tweet features for our FCN can be derived from previously learned fake news detection models or by concatenating extracted image and text features from CLIP.

\subsection{Label Propagation}
The straightforward approach for making a prediction is to use a Multi-layer Perceptron (MLP) to compute a two-dimensional vector of logits. Since the training and test sets for fake news detection may belong to different events, this can easily lead to overfitting. Integrating the predictions of relevant tweets to make a final decision facilitates coherent predictions across the entire set of tweets and mitigates overfitting. Simultaneously, edges in $\mathcal{G}$ can potentially represent positive and negative correlations. We thus design a label propagation network (LPN). This network smooths the labels considering both positive and negative correlations. Specifically, LPN comprises stacked bi-directional label attention layers. 

Let $\mathbf{x}^L_i$ denote the contextualized features from the final layer FPN. This feature vector is fed into an MLP to generate a two-dimensional vector of logits. Subsequently, a softmax function is applied to convert the logits into class probabilities $\mathbf{\tilde{y}}_i = (\tilde{y}^\text{r}_i, \tilde{y}^\text{f}_i)$, where $\tilde{y}^\text{r}_i$ and $\tilde{y}^\text{f}_i$ represent the probabilities of tweet $d_i$ being real or fake, respectively.

We use $L'$ label attention layers in our network. Given label $\hat{\mathbf{y}}_i^{l} \in [0,1]^2$ from the $l$-th label attention layer ($l = 1,\dots,L'-1$) and $\hat{\mathbf{y}}_i^{1} = \tilde{\mathbf{y}}_i$. The update for the $(l+1)$-th label $\hat{\mathbf{y}}_i^{l+1} \in [0,1]^2$ is given by: 
\begin{equation}
\hat{\mathbf{y}}_i^{l+1} = \mbox{softmax}\left(\hat{\mathbf{y}}_i^{l} + \sum_{j \in \mathcal{N}_i} \boldsymbol{\alpha}_{ij} \odot \hat{\mathbf{y}}_j^{l}\right),
\label{eq:lpa}
\end{equation}
where $\odot$ is element-wise multiplication and $\boldsymbol{\alpha}_{ij} = (\alpha_{ij}^\text{r}, \alpha_{ij}^\text{f}) \in [-1,1]^2$ is the bi-directional label attention weights between tweets $i$ and $j$. These weights are based on the contextualized features:
\begin{equation}
\boldsymbol{\alpha}_{ij} = \text{tanh}(A[V\mathbf{x}_i^L, V\mathbf{x}_j^L]),
\label{eq:alpha}
\end{equation}
where $[\cdot,\cdot]$ is concatenation and $\text{tanh}$ is hyperbolic tangent. $A \in \mathbb{R}^{2 \times 2d}$ and $V \in \mathbb{R}^{d \times d}$ are learnable weight matrices.

This bi-directional attention effectively captures both positive and negative correlations between label probabilities of neighboring tweets with ${\boldsymbol\alpha}_{ij}$. For example, if $\alpha_{ij}^\text{r}$ is negative, then tweet $d_j$'s probability of being real in a certain layer decreases the $d_i$'s probability of being real in the next layer. This flexibility allows for capturing various relationships between tweets to improve reliability and accuracy.

\begin{table*}[ht]
\centering
\Large
\caption{The performance of the state-of-the-art baselines is compared with the performance after integrating our FCN-LP with them, evaluated on the Twitter, PHEME, and Weibo datasets.}
\resizebox{\textwidth}{!}{
\renewcommand{\arraystretch}{1.1}
\begin{tabular}{lcccccccccccc} 
\toprule[2pt]
\multirow{2}{*}{Method}                       & \multicolumn{4}{c}{\textbf{Twitter}}                               & \multicolumn{4}{c}{\textbf{PHEME}}                                & \multicolumn{4}{c}{\textbf{Weibo}}                                  \\ 
\cmidrule(r){2-5}
\cmidrule(r){6-9}
\cmidrule(r){10-13}
                                              & Accuracy        & Precision      & Recall         & F1             & Accuracy       & Precision      & Recall         & F1             & Accuracy        & Precision      & Recall         & F1              \\ 
\midrule[0.5pt]
\textbf{EANN}              & 71.53$\pm$0.91  & 71.38$\pm$1.23 & 63.82$\pm$2.11 & 68.91$\pm$1.58 & 70.17$\pm$0.79 & 71.28$\pm$1.32 & 67.36$\pm$2.17 & 69.10$\pm$1.83 & 79.18$\pm$0.76  & 80.31$\pm$1.23 & 78.52$\pm$0.32 & 79.44$\pm$2.13  \\
\textbf{+ FCN-LP}                             & 74.21$\pm$2.12  & 73.52$\pm$0.51 & 67.11$\pm$0.65 & 71.33$\pm$2.14 & 74.22$\pm$1.16 & 75.11$\pm$2.41 & 70.41$\pm$1.52 & 72.33$\pm$1.17 & 82.23$\pm$1.53  & 83.56$\pm$0.77 & 80.98$\pm$1.56 & 81.63$\pm$1.77  \\
\midrule[0.5pt]
\textbf{SpotFake}  & 77.16$\pm$1.57  & 75.32$\pm$1.14 & 87.83$\pm$0.63 & 85.14$\pm$0.07 & 81.37$\pm$2.38 & 79.53$\pm$2.27 & 81.22$\pm$2.43 & 79.43$\pm$0.75 & 86.39$\pm$2.51  & 86.12$\pm$0.53 & 87.17$\pm$2.63 & 83.22$\pm$1.41  \\
\textbf{+ FCN-LP}                             & 79.46$\pm$2.61  & 78.21$\pm$1.51 & 88.41$\pm$1.02 & 86.53$\pm$2.15 & 83.62$\pm$0.87 & 81.41$\pm$0.24 & 83.17$\pm$1.08 & 81.76$\pm$0.47 & 86.88$\pm$0.56  & 87.65$\pm$1.78 & 87.44$\pm$0.21 & 85.53$\pm$1.04  \\
\midrule[0.5pt]
\textbf{MVAE}          & 74.56$\pm$1.58  & 80.15$\pm$2.69 & 76.34$\pm$0.83 & 81.57$\pm$1.98 & 77.83$\pm$1.27 & 73.82$\pm$2.05 & 73.45$\pm$2.62 & 72.21$\pm$0.54 & 71.86$\pm$0.25  & 70.32$\pm$0.69 & 70.32$\pm$2.84 & 70.53$\pm$1.60  \\
\textbf{+ FCN-LP}                             & 76.37$\pm$2.57  & 81.18$\pm$0.97 & 79.16$\pm$1.51 & 82.91$\pm$2.23 & 79.06$\pm$2.46 & 76.31$\pm$1.56 & 75.93$\pm$3.00 & 73.88$\pm$1.26 & 73.54$\pm$1.96  & 72.45$\pm$2.37 & 74.65$\pm$0.97 & 74.01$\pm$2.01  \\
\midrule[0.5pt]
\textbf{SAFE}      & 76.66$\pm$3.00  & 76.32$\pm$1.94 & 75.41$\pm$2.12 & 76.37$\pm$2.85 & 81.25$\pm$1.34 & 79.92$\pm$2.76 & 79.11$\pm$1.45 & 79.69$\pm$2.67 & 84.91$\pm$2.10  & 83.81$\pm$1.58 & 82.19$\pm$1.16 & 83.01$\pm$1.70  \\
\textbf{+ FCN-LP}                             & 77.94$\pm$0.99  & 79.09$\pm$0.45 & 76.73$\pm$1.63 & 77.55$\pm$2.32 & 83.00$\pm$2.60 & 82.22$\pm$1.62 & 81.44$\pm$0.46 & 82.38$\pm$0.91 & 87.14$\pm$2.89  & 85.30$\pm$2.59 & 85.05$\pm$0.92 & 84.53$\pm$0.64  \\
\midrule[0.5pt]
\textbf{MCAN}       & 80.91$\pm$2.33  & 82.68$\pm$2.48 & 76.67$\pm$0.94 & 82.26$\pm$1.32 & 80.74$\pm$1.89 & 79.21$\pm$2.23 & 79.64$\pm$1.53 & 80.15$\pm$0.86 & 89.41$\pm$2.91 & 90.35$\pm$2.17  & 88.51$\pm$1.78 & 90.01$\pm$1.57  \\                                          
\textbf{+ FCN-LP}                             & 81.98$\pm$1.90  & 83.51$\pm$1.11 & 77.24$\pm$2.21 & 83.19$\pm$0.98 & 81.30$\pm$1.20 & 80.49$\pm$1.91 & 80.50$\pm$1.72 & 81.03$\pm$2.13 & \pmb{91.52$\pm$2.67} & 91.56$\pm$2.43  & 89.20$\pm$1.11 & \pmb{91.40$\pm$1.91}  \\
\midrule[0.5pt]
\textbf{HMCAN}    & 83.91$\pm$1.49  & 81.68$\pm$2.08 & 84.67$\pm$1.21 & 82.57$\pm$1.62 & 86.36$\pm$1.83 & 83.18$\pm$1.41 & 83.81$\pm$2.51 & 83.45$\pm$1.07 & 88.51$\pm$2.90 & 92.04$\pm$2.97  & 84.57$\pm$1.77 & 88.11$\pm$1.14  \\    
\textbf{+ FCN-LP}                             & 84.57$\pm$1.62  & 83.58$\pm$1.66 & 85.22$\pm$2.06 & 84.04$\pm$0.82 & \pmb{87.25$\pm$1.18} & 84.77$\pm$1.91 & 84.78$\pm$1.95 & 84.50$\pm$0.85 & 91.15$\pm$2.82 & \pmb{94.01$\pm$2.56}  & 86.55$\pm$1.54 & 89.13$\pm$1.77  \\
\midrule[0.5pt]
\textbf{CLIP}      & 80.93$\pm$1.27  & 75.71$\pm$1.68 & 92.07$\pm$0.85 & 85.45$\pm$0.47 & 80.36$\pm$1.93 & 84.43$\pm$1.27 & 89.12$\pm$0.12 & 86.71$\pm$1.88 & 82.92$\pm$0.54 & 83.17$\pm$1.00 & 88.45$\pm$2.13 & 86.74$\pm$0.41   \\
\textbf{+ FCN-LP}                             & \pmb{89.00$\pm$1.90}  & \pmb{83.85$\pm$1.29} & \pmb{97.63$\pm$1.67} & \pmb{91.22$\pm$1.11} & 84.65$\pm$0.49 & \pmb{88.79$\pm$0.92} & \pmb{89.81$\pm$1.26} & \pmb{89.30$\pm$0.53} & 84.47$\pm$1.66 & 88.41$\pm$0.26 & \pmb{91.18$\pm$0.69} & 89.78$\pm$0.84   \\
\bottomrule[1.5pt]
\end{tabular}}
\label{comparison}
\end{table*}

\subsection{Classifcation with Domain Generalization}
Our approach consists of FCN and LPN. One of the main challenges in training a model for fake news detection is overfitting to the training data, resulting in poor performance on unseen tweets, particularly about new events. For generalization, it is crucial to ensure a consistent distribution of features of both training and unseen data. We thus divided the training dataset into two disjoint subsets $\mathcal{D}_\text{s}$ and $\mathcal{D}_\text{u}$ with seen and unseen tweets, respectively. The seen subset is used to train both FCN and LPN, whereas the unseen subset is only for updating FCN, so that the distribution of unseen tweets' features is consistent with seen ones. 

Specifically, we train FCN and LPN by minimizing the cross-entropy loss of predicted labels over $\mathcal{D}_\text{s}$:
\begin{equation}
\mathcal{L}_{\text{FCN}} = -\sum_{d_i \in \mathcal{D}_\text{s}} \sum_{k\in\{\text{r},\text{f}\}} y_i^k \log \tilde{y}_i^k(\mathbf{x}_i^L),
\label{eq:gcnloss}
\end{equation}
where $y_i^k$ is the ground-truth label for tweet $d_i$ and $\tilde{y}_i^k(\mathbf{x}_i^L)$ is the corresponding prediction.\footnote{We explicify that $\hat{y}_i^k$ is a function of $\mathbf{x}_i^L$.} This loss function will maximize the probability that each seen node in the training set is correctly labeled, thus increasing the inter-class variation in node feature $\mathbf{x}_i^L$. 

We also apply the cross-entropy loss to train LPN:
\begin{equation}
\mathcal{L}_{\text{LPN}} = -\sum_{d_i \in \mathcal{D}_\text{s}} \sum_{k\in\{\text{r},\text{f}\}} y_i^{k} \log \hat{y}_i^{L'k}(\mathbf{x}_i^L, \tilde{\mathbf{y}}_i),
\label{eq:lpnloss}
\end{equation}
where $\hat{y}_i^{L'k}$ is the output of the last ($L'$)-th layer of LPN. Based on this training, FCN learns initial classification prediction labels from features of tweets in $\mathcal{D}_\text{s}$, and LPN learns to efficiently aggregate neighboring tweets' labels to take their correlations into account.

We next consider generalizing the model to unseen tweets. A reasonable generalization is that guarantees accurate classification on the seen tweets while regularizing the model to maintain a consistent distribution on the unseen tweets. Assuming we have two sets $\mathcal{P}$ and $\mathcal{Q}$ of features, we can use Maximum Mean Discrepancy (MMD) to compare the differences between the two distributions in the feature space as:
\begin{equation}
\text{MMD}(\mathcal{P}, \mathcal{Q}) = \left\| \frac{1}{|\mathcal{P}|}\sum_{\mathbf{p} \in \mathcal{P}} \mathbf{p} - \frac{1}{|\mathcal{Q}|}\sum_{\mathbf{q} \in \mathcal{Q}} \mathbf{q} \right\|_2^2.
\end{equation}

To ensure intra-class consistency and inter-class distinguishability of features, we compute the MMD separately for the real and fake features of tweets between seen and unseen sets and use a sum of these losses as the overall loss. Specifically, given the seen feature sets with real and fake labels $\mathcal{X}_{\text{s}}^{\text{r}}$ and $\mathcal{X}_{\text{s}}^{\text{f}}$, as well as the unseen feature sets with real and fake labels $\mathcal{X}_{\text{u}}^{\text{r}}$ and $\mathcal{X}_{\text{u}}^{\text{f}}$, we can define the overall loss as follows:
\begin{equation}
\mathcal{L}_{\text{MMD}} = 
\text{MMD}(
  \mathcal{X}_{\text{s}}^{\text{r}}, 
  \mathcal{X}_{\text{u}}^{\text{r}}) 
+ \text{MMD}(
  \mathcal{X}_{\text{s}}^{\text{f}},
  \mathcal{X}_{\text{u}}^{\text{f}}
).
\end{equation}
By minimizing the $\mathcal{L}_{\text{MMD}}$ loss, we can ensure the consistency of features between seen and unseen tweets in the feature space, thereby improving the model's generalization ability. Finally, we combine the three losses ($\mathcal{L}_{\text{FCN}}$, $\mathcal{L}_{\text{LPN}}$, and $\mathcal{L}_{\text{MMD}}$) and train the entire model end-to-end with the following loss function:
\begin{equation}
\mathcal{L}_{all} = \mathcal{L}_{\text{FCN}} +\lambda\mathcal{L}_{\text{LPN}} + \mu\mathcal{L}_{\text{MMD}}
\label{eq:allloss}
\end{equation}
where $\lambda$ and $\mu$ are the balancing hyper-parameters.

\begin{table*}[ht]
\centering
\small
\caption{Performance comparison of FCN-LP with its variants using CLIP features on the Twitter, PHEME, and Weibo datasets. }
\resizebox{\textwidth}{!}{
\begin{tabular}{lcccccccccccccccc} 
\toprule[1pt]
 &\multirow{2}{*}{\small{\textbf{FCN}}}      &\multirow{2}{*}{\small{\textbf{LPN}}}&\multirow{2}{*}{\small{\textbf{LPN}-$\boldsymbol{\alpha}$}}&\multirow{2}{*}{\small{$\mathcal{L}_{\text{MMD}}$}}& \multicolumn{4}{c}{\textbf{Twitter}}                               & \multicolumn{4}{c}{\textbf{PHEME}}                                & \multicolumn{4}{c}{\textbf{Weibo}}                                  \\ 
\cmidrule(r){6-9}
\cmidrule(r){10-13}
\cmidrule(r){14-17}
  &  &   &    &     & Acc.        & Prec.      & \small{Recall}         & F1             & Acc.        & Prec.      & \small{Recall}   & F1             & Acc.        & Prec.      & \small{Recall}  & F1              \\ 
\midrule[0.5pt]
(i) &  &   &    &     & 80.93 & 75.71 & 92.07 & 85.45 & 80.36 & 84.43 & 89.12 & 86.71 & 82.92 & 83.17 & 88.45 & 86.74 \\
(ii)&$\checkmark$ &    &    &                  & 82.17 & 76.54 & 92.82 & 86.63 & 80.47 & 85.92 & 89.35 & 86.85 & 83.15 & 84.26 & 88.94 & 86.92 \\
(iii)&$\checkmark$& $\checkmark$&    &          & 87.62 & 81.58 & 96.17 & 89.82 & 83.27 & 87.28 & 89.76 & 88.75 & 84.22 & 87.52 & 90.21 & 88.77 \\
(iv)&$\checkmark$& & $\checkmark$&$\checkmark$ & 86.82 & 80.87 & 96.01 & 87.93 & 82.15 & 86.34 & 89.71 & 88.12 & 83.91 & 86.25 & 89.77 & 88.21 \\
(v)&$\checkmark$&  $\checkmark$ &    &    $\checkmark$ &  \pmb{89.00}  & \pmb{83.85} & \pmb{97.63} & \pmb{91.22} & \pmb{84.65} & \pmb{88.79} & \pmb{89.81} & \pmb{89.30} & \pmb{84.47} & \pmb{88.41} & \pmb{91.18} & \pmb{89.78}   \\
\bottomrule[1pt]
\end{tabular}}
\label{ablation}
\end{table*}

\section{Experiment}
\subsection{Datesets}
We evaluated our model on three real social media datasets from Twitter, PHEME, and Weibo.

\textbf{Twitter} \cite{boididou2015verifying}: The Twitter dataset, part of the MediaEval Verifying Multimedia Use benchmark \cite{boididou2015verifying}, is used for detecting fake content on social media and contains approximately 17,000 unique tweets related to various events. Each tweet includes textual content, an attached image or video, and additional social context, with labels on fake/real and the event. We removed tweets with videos. After removal, the dataset is divided into a training set (9,000 fake and 6,000 real news tweets) and a test set (2,000 tweets), ensuring no overlapping events. There are 15 events in the training set, and we randomly select tweets related to 11 events as seen tweets and the remaining as unseen tweets.

\textbf{PHEME} \cite{zubiaga2017exploiting}: The PHEME dataset comprises tweets from Twitter, focusing on five breaking news events. Each event contains a set of posts, including a large number of texts and images, each pair of which also comes with labels on fake/real and the event. We randomly select three events and extract related tweets for training, where the tweets on randomly selected two events are used as seen tweets while the ones on the other event are used as unseen tweets. The tweets on events other than these three are used for evaluation.

\textbf{Weibo} \cite{jin2017multimodal}: The Weibo dataset is collected from a Chinese microblogging service, Sina Weibo. The real news is from authoritative Chinese news sources, such as Xinhua News Agency, while the fake news is collected from Weibo between May 2012 and January 2016, verified by Weibo's official rumor debunking system. We remove tweets without any text or image. This dataset does not provide event labels, so we randomly split the dataset into seen, unseen, and testing sets at 6:2:2. Each sample in the dataset contains tweet ID, text, image, and real/fake label.

\subsection{Baselines} \label{sec:baseline}

To verify the validity of our model, we selected several state-of-the-art multimodal methods for comparison.

\textbf{EANN} \cite{wang2018eann} first extracts textual and visual features separately using pre-trained models and combines these features to feed into a fake news detector. To make the features invariant to events, they simultaneously train an event classifier and the detector in an adversarial manner so that the classifier fails.

\textbf{SpotFake} \cite{singhal2019spotfake} detects fake news by using a simple concatenation of features extracted from pre-trained text and image models.

\textbf{MVAE} \cite{khattar2019mvae} utilizes a bimodal variational autoencoder to learn a shared representation of tweets, coupled with a binary classifier to detect fake news.

\textbf{SAFE} \cite{zhou2020similarity} jointly exploits multimodal features and cross-modal similarity for learning tweet features.

\textbf{MCAN} \cite{wu2021multimodal} leverages co-attention layers to learn the inter-modal relationships between textual and visual features. The enhanced features are fused for fake news detection.

\textbf{HMCAN} \cite{qian2021hierarchical} utilizes a multi-modal contextual attention network to fuse inter-modality and intra-modality relationships. A hierarchical encoding network captures rich hierarchical semantics, improving fake news detection.

\textbf{CLIP} \cite{radford2021learning} exhibits strong zero-shot capabilities for several vision-and-language tasks. We train a fake news detector with a 2-layer MLP that takes as input a concatenation of image and text features extracted by CLIP.

\subsection{Implementation Details} 
We implement our FCN-LP with Pytorch and PyG (PyTorch Geometric).\footnote{The code is available at \url{https://github.com/zhaowanqing/FCN-LP}.} For all the experiments, we used NVIDIA GeForce GTX 3090 GPU with 24GB memory. For constructing the cross-modal tweet graph, we choose a similarity threshold $\tau= 0.95$ on the Twitter dataset and $\tau= 0.91$ on the PHEME and Weibo datasets. The balancing hyper-parameters $\lambda$ and $\mu$ are both set to 1. These choices are based on our parameter sensitivity analysis. FCN-LP has four graph convolutional layers and two label attention layers, which are trained using AdamW with the learning rate of $10^{-4}$ for 500 epochs. We train the model 5 times and report the average and standard deviation of accuracy, precision, recall, and F1 Score.

\subsection{Performance Comparison} 
One advantage of FCN-LP is that it can be cascaded to any fake news detector to improve its performance. For this, we first train a detector over a fake news dataset. Then, we extract tweet features from the layer preceding the prediction head, which will serve as node features of FCN-LP. For the CLIP baseline, we directly concatenate the image and text features by CLIP as node features. These features are fixed while training FCN-LP. 

The results are shown in Table \ref{comparison}. We observe that FCN-LP can boost the performance of the existing multimodal fake news detectors in all metrics, ranging from 0.57\% to 8.07\%. This may mean that interactions among features and modeling positive and negative correlations via a graph based on multimodal similarities facilitate the detection. In particular, the detection based on CLIP features shows superior performance on Twitter and PHEME datasets. This can be attributed to CLIP being pre-trained on a larger multimodal dataset, resulting in features with better generalization when applied to the test set. After training on a fake news dataset, tweet features extracted from other methods may suffer from potential overfitting. This leads to limited performance improvement by FCN-LP compared to the CLIP-based baseline. On the other hand, since CLIP's pre-training dataset is mainly from an English corpus, it does not show strong performance in the Weibo dataset, which is dominated by Chinese text. MCAN and HMCAN employ Chinese BERT to extract text features. The enhancement in their performance on the Weibo dataset can probably be attributed to this language model.

\subsection{Ablation Study}

Several design choices are ablated within FCN-LP to comprehend their importance better. All ablation experiments uses CLIP's textual and visual features. We evaluated five settings: (i) CLIP baseline (as in Section \ref{sec:baseline}). (ii) FCN-only that solely utilizes FCN to contextualize the features and feeds them into a 2-layer MLP for fake new detection. (iii) FCN-LPN without MMD that integrates both FCN and LPN but omits the Maximum Mean Discrepancy (MMD) loss. All tweets in the training set are used as seen data for training. (iv) LPN-$\alpha$: To evaluate the impact of bi-directional label attention, we modified $\boldsymbol{\alpha}_{ij}$ in LPN to a 1-dimensional attention weight, ranging from $[0,1]$. This variant highlights the impact of using positive and negative correlations in label propagation. (v) FCN-LP. Due to MMD, these five variants are exposed to the training set differently. (i)--(iv) use the entire training set for supervised training, whereas (v) uses the seen subset to train FCN in a supervised manner and both the seen and unseen subsets for MMD. This means that FCN-LP uses fewer samples for supervised training. We consider that this difference advantages (i)--(iv) but not our full model. 

\emph{1) Effects of feature contextualization:} From the comparison between (i) and (ii) in Table~\ref{ablation}, we can observe that feature contextualization by neighboring tweets is helpful. Compared to detectors solely based on individual tweets, incorporating additional context can enhance the understanding of a certain tweet's content and allow for comparison.

\emph{2) Effects of label propagation:} As illustrated in Table~\ref{ablation}, the average accuracy of (iii) exhibited an improvement compared to (ii) by 5.45\%, 2.8\%, and 1.07\% on the Twitter, PHEME, and Weibo datasets, respectively. Since the training set and test set of Twitter and PHEME are related to different events, the risk of overfitting is more obvious. The label propagation can make a final decision by combining the prediction of relevant tweets, achieving a more cohesive and accurate analysis across the entire tweet set, and effectively reducing the possibility of overfitting. LPN offers more significant improvements in the Twitter and PHEME datasets, potentially indicating that our method robustifies detection for new events.

\emph{3) Effects of bi-directional label attention:} Variant (iv) shows the effectiveness of bi-directional label attention. Compared to (v) FCN-LP, (iv) shows a significant performance deterioration. The possible reason is that there is a potentially positive and negative correlation between relevant tweets and single-valued attention experiences difficulty in discerning these relationships, leading to wrong predictions. Bi-directional label attention, in contrast, effectively captures such various relationships and assigns appropriate weights to predictions.

\emph{4) Effects of domain generalization:} We can verify the effect of the domain generalization loss $\mathcal{L}_{\text{MMD}}$ by comparing  (iii) and (v). As seen in Table~\ref{ablation}, the domain generalization loss gives an obvious boost in all datasets. This implies that constraining the features to distribute a certain (the seen subset's) volume via $\mathcal{L}_{\text{MMD}}$ would be beneficial in classifying new tweets that may be related to unseen events. This may mean that $\mathcal{L}_{\text{MMD}}$ consequently tries to make the features invariant to events.

\begin{figure}[t]
\includegraphics[width=0.7\columnwidth]{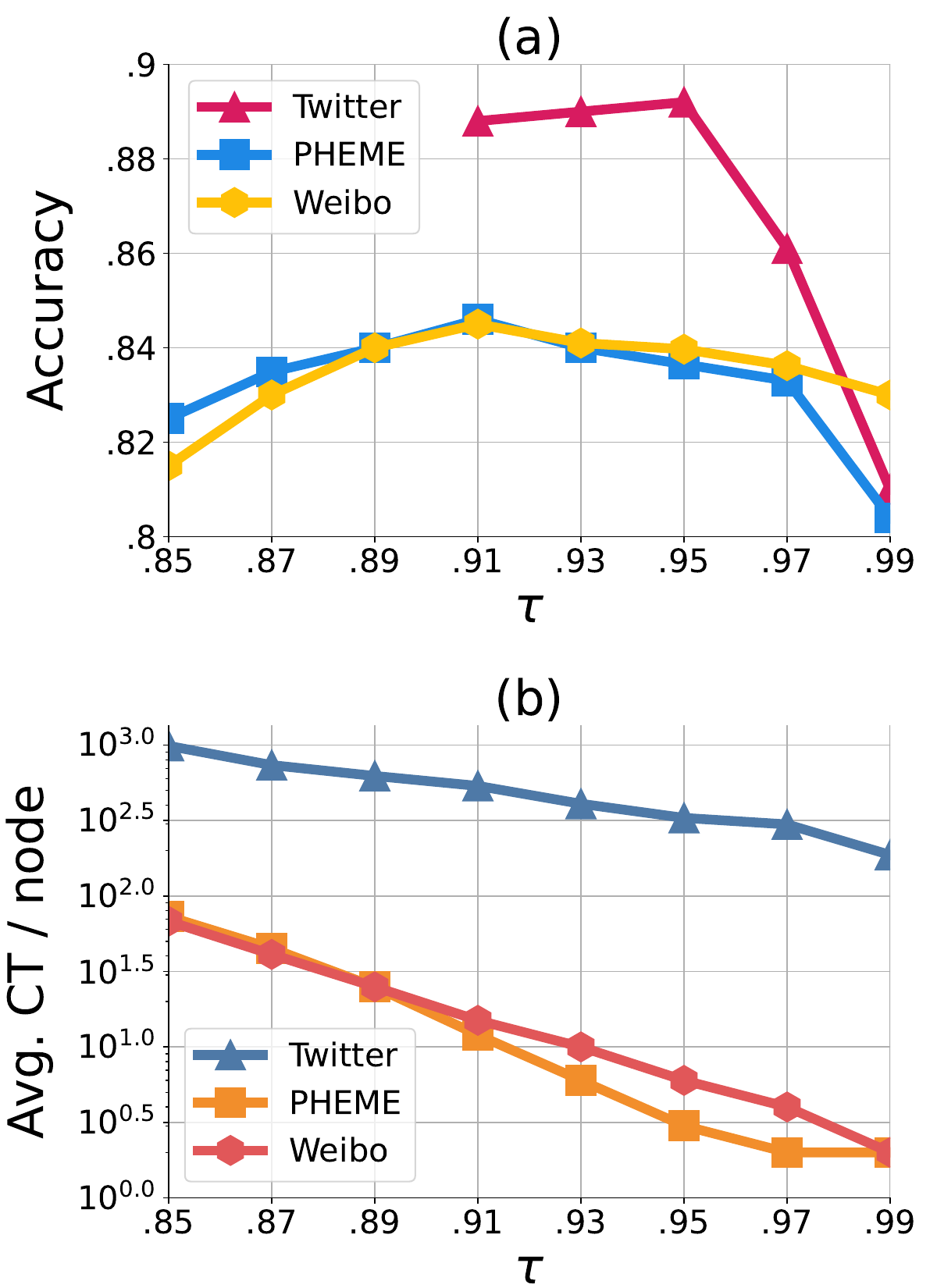}
\centering
\caption{\label{fig:para-ana} Parameter sensitivity analysis for similarity threshold $\tau$: (a) accuracy curves with different similarity thresholds; (b) the average number of connected (relevant) tweets per node (Avg. CT / node). }
\end{figure}

\begin{figure}[t]
\includegraphics[width=0.8\columnwidth]{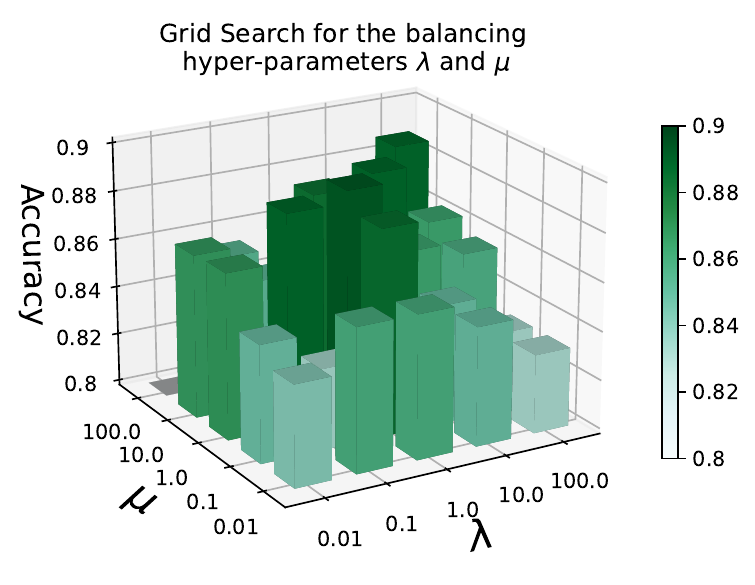}
\centering
\caption{\label{fig:grid} Accuracy values obtained for various loss balancing hyper-parameters $\lambda$ and $\mu$ on the Twitter dataset.}
\end{figure}

\begin{figure}[t]
\includegraphics[width=0.8\columnwidth]{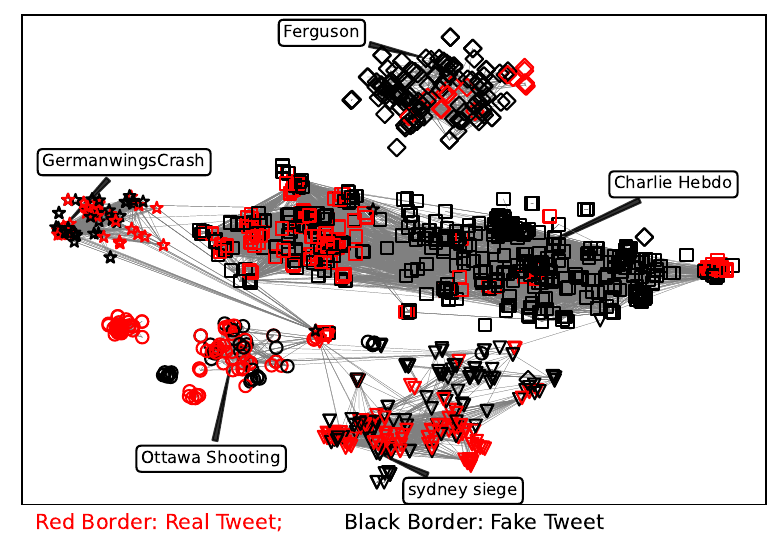}
\centering
\caption{\label{fig:tweetgraph} 2D t-SNE plot of the cross-modal tweet graph constructed on the PHEME dataset. The various markers signify distinct events within the dataset. A red border indicates the tweet is real, while a black border denotes a fake tweet.}
\end{figure}

\begin{figure}[t]
\includegraphics[width=\columnwidth]{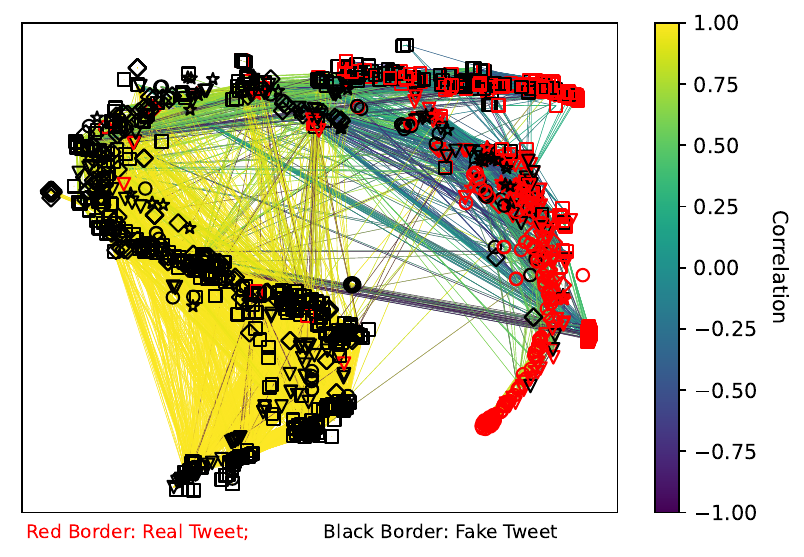}
\centering
\caption{\label{fig:FCNgraph} Visualization of the learned FCN features constructed on the PHEME dataset. The color lines indicate bi-directional attention values on connected tweets. The yellow line indicates a positive correlation, and the blue line indicates a negative correlation.}
\end{figure}

\begin{figure*}[t]
\includegraphics[width=\textwidth]{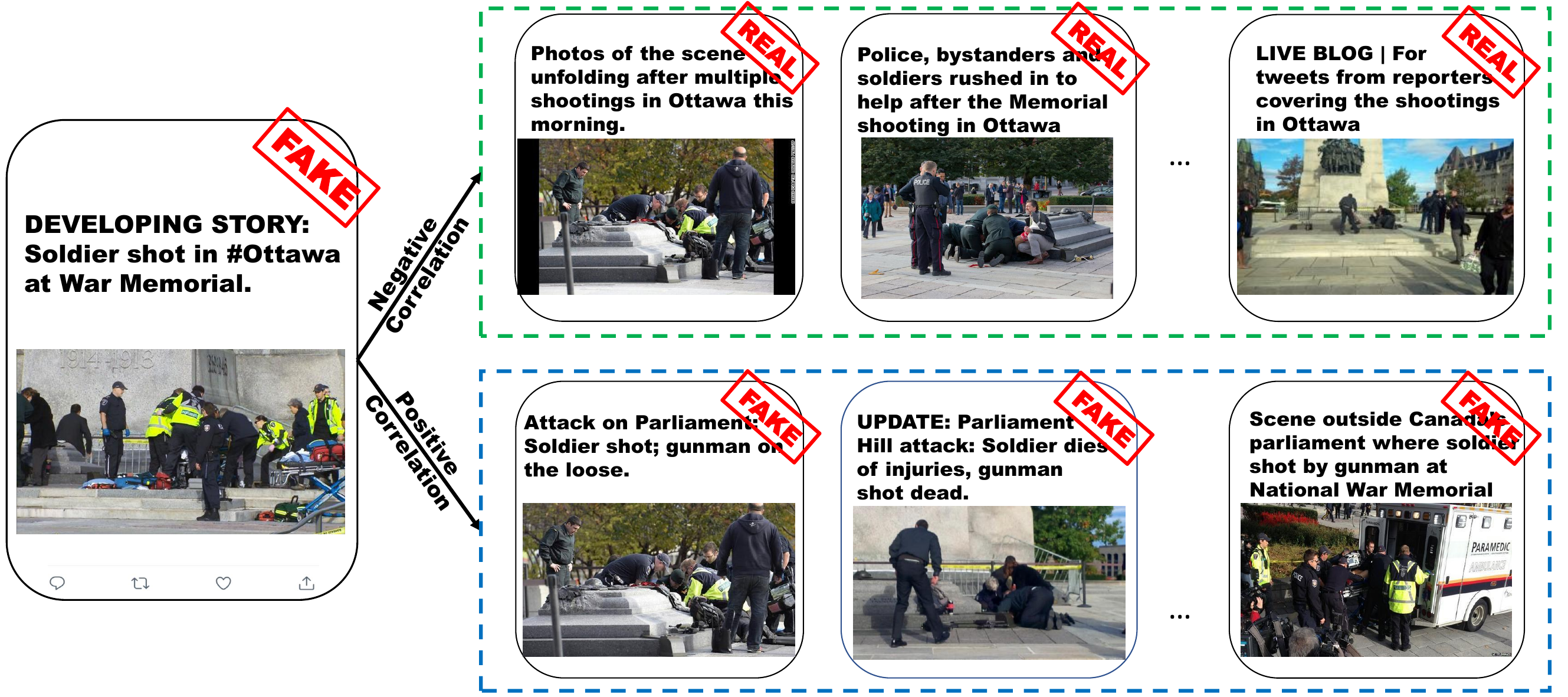}
\centering
\caption{\label{fig:cases}Example of a tweet query: leveraging our method, we can not only detect fake news but also present associated real and fake tweets based on correlations. This approach offers interpretable classification results and assists users in discerning the fact behind the information.} 
\end{figure*}

\subsection{Parameters Sensitivity Analysis} 
We study the sensitivities of 1) $\tau$, the similarity threshold in constructing the cross-modal tweet graph, and 2) $\lambda$ and $\mu$, the balancing hyper-parameters in the overall loss function. All parameter sensitivity analysis experiments were done using the CLIP model.

The similarity threshold $\tau$ dictates the strength of connections between tweets based on their content similarity. A higher threshold value ensures more strict and precise connections. Conversely, a lower threshold value allows for a greater number of connections but may introduce noise in the form of irrelevant or loosely related tweets. We vary the value of $\tau$ from $0.85$ to $0.99$. We report the accuracy with different $\tau$ in Figure~\ref{fig:para-ana}(a) and the average number of connections per node in Figure~\ref{fig:para-ana}(b). Since a large number of tweets correspond to the same image in the Twitter dataset, more tweets are connected to each node than in the PHEME and Weibo datasets. Too many connections led to out-of-memory on the Twitter dataset, so we excluded the cases where $\tau < 0.91$. We finally chose $\tau = 0.95$ on the Twitter dataset and 0.91 on the PHEME and Weibo datasets, which better balances between the accuracy and complexity.

To analyze the balancing hyper-parameters in the loss function, we change the values of $\lambda$ and $\mu$ between 0.01 and 100. We performed a grid search over this range and chose the best hyper-parameters. A bar plot of the validation accuracy on the Twitter dataset for various combinations of $\lambda$ and $\mu$ is given in Figure~\ref{fig:grid}. It can be noticed that similar weights among the three loss functions show better performance. Therefore the balancing hyper-parameters in our approach can be set as $\lambda=\mu=1$. A similar behavior was noticed on the PHEME and Weibo datasets as well.

\subsection{Visualization}
We visualize the cross-modal tweet graph constructed on the PHEME dataset in Figure~\ref{fig:tweetgraph}. We use t-SNE \cite{van2008visualizing} to reduce the original tweet features to 2D and sample 10\% for visualization. As seen in Figure~\ref{fig:tweetgraph}, CLIP features separate events at divisible distances, and neighbors usually have a connection. This is consistent with our estimation and indicates that cross-modal tweet graphs are meaningfully used for aggregating features and label propagation. 

In order to further explore the learned FCN feature and label propagation between tweets, we visualize the learned FCN features in Figure~\ref{fig:FCNgraph} using the same way as visualizing the cross-modal tweet graph, with the addition of color lines to indicate attention values on labels. We can see that the learned FCN features show a clear distinction between real and fake tweets. There are some connections in the graph between real and fake tweets, but our bi-directional attention gives them a negative correlation and, therefore indirectly contributes to a performance boost.

This cross-modal tweet graph with bi-directional attention can potentially provide interpretability for classification results and even reveal the fact about the query tweet. For example, as shown in Figure~\ref{fig:cases}, for a tweet predicted as fake, we can simultaneously provide associated real and fake tweets along with their correlation, serving as fact-checked information to improve users’ consciousness of fake news. As seen in the example, the relevant matching real event is a shooting incident in Ottawa. However, it does not explicitly mention the soldier being shot in real tweets, and the fake tweet could potentially lead to misinterpretation.

\section{Conclusions} 
This paper presents a method for detecting fake news on social media by leveraging cross-modal tweet content. Our method constructs a cross-modal tweet graph using CLIP, which allows us to extract potential connections and reveal the events behind them to improve detection performance. We also designed FCN-LP to learn the signed attention of the edges, which generates positive or negative correlations applied to the related tweet labels. Our method can generalize even to tweets on unseen events. Our label propagation based on a constructed cross-modal tweet graph is portable. We can put it on top of existing methods and use their own predictions to improve the detection performance. Our domain generalization loss also improves the model's ability to detect unseen events, making it more robust in real-world scenarios. One potential avenue for future research is exploring the use of pre-trained large language models, such as GPT-3 or even GPT-4, for fake news detection. These models have shown impressive performance on a range of natural language processing tasks. They may be able to reveal biases or distortions in the text that are indicative of fake news. 
\section*{Acknowledgments}
This research was supported by the National Natural Science Foundation of China (Grant No. 62273275), the China Scholarship Council and the JST CREST Grant No.~JPMJCR20D3.
%%
%% The next two lines define the bibliography style to be used, and
%% the bibliography file.
\bibliographystyle{ACM-Reference-Format}
\balance

%%
%% If your work has an appendix, this is the place to put it.

\end{document}